\documentclass[twocolumn,secnumarabic,amssymb, nobibnotes, aps,notitlepage]{revtex4-1}
\usepackage{graphicx}
\usepackage{hyperref}
\begin{document} 
\title{Challenges to the Good-Walker paradigm in 
coherent and incoherent photoproduction}

\author{Spencer R. Klein}%
\email[]{srklein@lbl.gov}
\affiliation{Nuclear Science Division, Lawrence Berkeley National Laboratory, 1 Cyclotron Road, Berkeley, CA 94720 USA}
\date{\today}%
\begin{abstract}

High-energy vector meson photoproduction is an important tool for studying the partonic structure of matter at low Bjorken$-x$.   In the Good-Walker (GW) paradigm, the cross-section $d\sigma/dt$ for coherent production of vector mesons or other final states, depends the average transverse distribution of gluons, while the incoherent cross-section depends on fluctuations in the nuclear structure, due to variations in nucleon positions, and/or gluonic hot spots.  However, predictions of the the GW paradigm seemingly conflict with data from multiple experiments which observe coherent production of vector mesons accompanied by nuclear excitation, or in peripheral relativistic heavy-ion collisions.  These data are consistent with a simpler, semi-classical approach. We will discuss this contradiction and explore how and why GW fails.  We will also contrast the significant differences in incoherent photoproduction on $^{197}$Au and $^{208}$Pb targets in the GW approach with the much smaller expected differences in their low$-x$ gluon content. 

\end{abstract}
\maketitle

\section{Introduction}

Vector meson photoproduction in ultra-peripheral collisions (UPCs)
\cite{Bertulani:2005ru,Baltz:2007kq,Contreras:2015dqa}
and at a future electron-ion collider (EIC) \cite{Accardi:2012qut,
AbdulKhalek:2021gbh}  is a key tool to probe nuclear structure at low Bjorken$-x$ \cite{Klein:2019qfb}. Exclusive vector meson production occurs when a photon fluctuates to a quark-antiquark pair ($q\overline q$, a virtual vector meson) which then scatters elastically from a target nucleus, emerging as a real vector meson.  

At high energies, the lifetime of the dipole is longer than the transit time through the nucleus, so it may be treated as remaining in a fixed configuration throughout the transit.  The elastic scattering may be described as occurring via the exchange of a Pomeron, which, to lowest order (LO), is composed of two gluons \cite{Donnachie:1988nj}.   The cross-sections to produce heavier vector mesons scale then with the square of the target gluon density.  Measurements of $J/\psi$ and $\psi'$ photoproduction cross-sections in UPCs at the LHC have been used to measure the gluon content of protons \cite{ALICE:2018oyo,LHCb:2018rcm} and lead nuclei \cite{ALICE:2021gpt,LHCb:2021bfl,CMS:2016itn}.  the data is consistent with moderate gluon shadowing for $x$ around $10^{-3}$ \cite{Guzey:2020ntc}. 
However, recent calculations of photoproduction at next-to-leading order show a more complicated theoretical picture, though \cite{Eskola:2022vpi}.   


Vector mesons measurements are also sensitive to the internal structure of the target \cite{Klein:2019qfb}.  In the Good-Walker (GW) picture, the differential coherent cross-section $d\sigma_{\rm coh}/dt$ ($t$ is the Mandelstam $t$) probes the transverse distribution of gluons in the target nucleus, as
was highlighted in the EIC White Paper \cite{Accardi:2012qut} and EIC Yellow Report \cite{AbdulKhalek:2021gbh}.  

The Yellow Report also considered incoherent photoproduction, which probes event-by-event fluctuations in the target configuration, including gluonic hot spots and nucleonic/subnucleonic target positions.  As the photon energy $k$ increases, the reaction probes gluons with lower $x$.  More and more gluonic hot spots appear, raising the incoherent cross-section \cite{Cepila:2018zky,Cepila:2016uku}.  As $k$ increases, the number of hot spots grows to eventually encompass the whole nucleus, turning it into a black disk.  Black disks don't fluctuate, and so, at sufficiently high $k$, the incoherent cross-section disappears.   Similar behavior is found with other approaches, such as a calculation using the JIMWLK equation, which found that, at HERA energies, the ratio of the incoherent to coherent cross-section dropped with increasing $k$ \cite{Mantysaari:2018zdd}.

After discussing the GW paradigm (Sec. II),confronting it with data (Sec. III), and giving a reason why it may fail (Sec. IV),  this paper will discuss other issues with the GW paradigm in Section V, before drawing sone conclusions in Sec. VI. 

\section{The Good-Walker Paradigm}

In 1960, Good and Walker studied high-energy diffractive dissociation \cite{Good:1960ba}.  If a single incident particle can be described as the sum of amplitudes for different states with different absorption cross-sections, then it may interact elastically (and coherently) with a target, and emerge as a different particle, as long as the incident and final state particles have the same quantum numbers, including spin and intrinsic angular momentum.  Good and Walker argued that, although the incident-particle plus target system could have orbital angular momentum, kinematic considerations make this unlikely \footnote{The only exception for nuclear targets is G-parity, because nuclei are not in fixed states of G-parity}.  

Mietenlin and Pumplin \cite{Miettinen:1978jb,Frankfurt:2022jns} extended the approach to include incoherent photoproduction.  The total cross-section for a diffractive process like photoproduction may be written \cite{Klein:2019qfb}
\begin{equation}
\frac{d\sigma_{\rm tot}}{dt} = \frac{1}{16\pi}\big<|A(K,\Omega)|^2\big>
\label{eq:tot}
\end{equation}
where $A(K,\Omega)$ is the amplitude for photoproduction. Here $K$ are the kinematic factors in the reaction, and $\Omega$ is the configuration of the target: the position of individual nucleons, and their parton configurations, including fluctuations.  The amplitudes are squared to get the cross-section for that configuration, and then averaged over configurations.   

In GW, in coherent interactions the target nucleus remains in its ground state. The  amplitudes are added, and
\begin{equation}
\frac{d\sigma_{\rm coh}}{dt} = \frac{1}{16\pi}\big|<A(K,\Omega)>\big|^2.
\label{eq:coh}
\end{equation}
This equation directly ties together two signatures: that the target remains in its ground state, and the coherent enhancement that is present when one adds amplitudes before squaring. 

The incoherent contribution is just the difference between the total and the coherent cross-sections: 
\begin{equation}
\frac{d\sigma_{\rm inc}}{dt} = \frac{1}{16\pi}\bigg(\big<|A(K,\Omega)|^2\big>-\big|\big<A(K,\Omega)\big>\big|^2\bigg).
\label{eq:inc}
\end{equation}
Event-by-event fluctuations in the nuclear configuration lead to differences between the two terms within the parentheses -  the sum of the squares minus the square of the sum.  Here, the momentum transfer $\sqrt{|t|}$ is related to the length scale for these fluctuations, but  one cannot use $d\sigma_{\rm inc}/dt$ to directly predict fluctuations on length scales $L=\hbar/\sqrt{|t|}$; Instead, $d\sigma_{\rm inc}/dt$ can be used for model-testing.  For example, HERA data on $J/\psi$ photoproduction on proton targets is consistent with strong geometric fluctuations of the proton target \cite{Mantysaari:2016ykx}. 

\section{Coherent photoproduction in non-exclusive reactions}

The GW paradigm is challenged by data showing that coherent photoproduction occurs even when the nuclear targets breaks up.  Here, coherent production is signaled by the presence of a peak in $d\sigma/dt$ (for $t<{\rm few}\ (\hbar/R_A)^2$), consistent with in-phase addition of the amplitudes to scatter from different nuclei \cite{STAR:2022wfe}.  

The first data came from the Solenoidal Tracker at RHIC (STAR) experiment at RHIC, which includes a central detector and two zero degree calorimeters (ZDCs) upstream and downstream of the interaction point.  The ZDCs detect neutrons from nuclear breakup.  STAR UPC data was collected with a trigger that requires one or more neutrons to be present in each ZDC.  This happens when both nuclei break up and emit neutrons.  STAR observed photoproduction of the $\rho$ plus direct $\pi^+\pi^-$ plus $\omega$ \cite{STAR:2002caw,STAR:2007elq,STAR:2017enh}, $\rho'$ \cite{STAR:2009giy} and the $J/\psi$ \cite{Adam:2019rxb}.  All exhibited coherent production.  ALICE has confirmed the coherent photoproduction of the $\rho$, in concert with breakup of one or both nuclei \cite{ALICE:2020ugp}.  
 
\begin{figure}
  \begin{center}
    \includegraphics[width=0.3\textwidth]{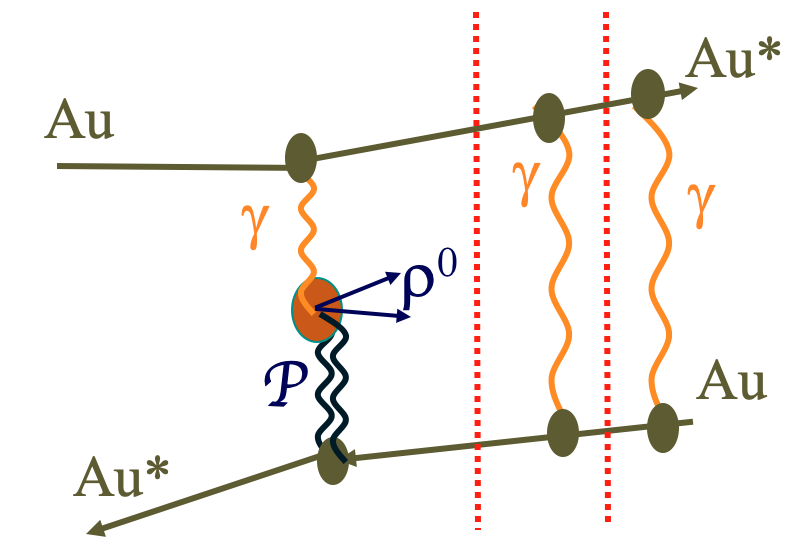}
    \caption{Schematic diagram of vector meson photoproduction accompanied by mutual Coulomb excitation.  The vertical dashed lines show that the photons are independent, sharing only a common impact parameter.}
    \label{fig:multiphoton}
  \end{center}
\end{figure}

 These cross-sections are consistent with the picture shown in Fig. \ref{fig:multiphoton}, where two (if one nucleus breaks up) or three (if both nuclei dissociate) photons are exchanged \cite{Baltz:2002pp}.  Each photon is emitted independently \cite{Gupta:1955zza} and does one thing: one coherently produces the vector meson, while the other photons break up one nucleus each. The photons are connected only through their common impact parameter \cite{Baur:2003ar}.   Although the photons act independently, they are still parts of the full reaction in Fig. \ref{fig:multiphoton}, and the intermediate ion lines may be off the mass shell.  This is likely not a significant effect, but it is present.  A competing, indistinguishable process becomes significant at larger $p_T$:  single photon exchange which both produces a vector meson and excites the target nucleus. 

The other class of data involves coherent $J/\psi$ photoproduction in peripheral collisions \cite{STAR:2019yox,ALICE:2022zso,LHCb:2021hoq}.   A large excess (over the hadroproduction expectations) of $J/\psi$ production  was observed at small $p_T$, with characteristics consistent with coherent photoproduction.  The excess was seen down to about 30\% centrality, {\it i. e.} for about 70\% of the hadronic cross-section, including collisions involving hundreds of participants, producing hundreds of final state particles.  This contrasts with the GW requirement that the target nucleus remain in its ground state.  

In contrast, these data are explained in a semi-classical approach to coherence, where, as long as one cannot tell which nucleon was struck, one adds the amplitudes for the photon/dipole to interact with the different nucleons in the target.
The cross-section to produce a vector meson on a nuclear target is
\begin{equation}
    \sigma = \big|\Sigma_i^N A_i \exp(i\vec{k}\cdot\vec{x}_i)\big|^2
    \label{eq:sigmasimp}
\end{equation}
where $i$ sums over the $N$ nucleons at positions $\vec{x}_i$, each with production amplitude $A_i$.  Here, $k$ is the momentum transfer from the target to the vector meson.  We take the $A_i$ to be identical, but that is not required for the argument.  

Equation \ref{eq:sigmasimp} leads to two different recoil spectra, corresponding roughly to the sizes of the entire nucleus and to that of an individual nucleon.   When $|\vec{k}| < \hbar/R_A$, the exponential is unity, and then $\sigma\approx N^2$; the cross-section is coherently enhanced.  For $|\vec{k}| \gg \hbar/R_A$, the phase of the exponential is random and $\sigma\approx N$; there is no coherent enhancement; the $p_T$ spectrum depends on the structure of the proton. These two regimes are observed in UPC data, with $p_T$ spectra in qualitative agreement with interactions from the entire nucleus and from individual nucleons \cite{STAR:2007elq}.

In contrast to GW, Eq. \ref{eq:sigmasimp} takes each nucleus as it appears, without reference to an average configuration.  One could generate new nuclear configurations for each interaction, but, for large nuclei, the incoherent and coherent cross-sections would not change significantly. 

The semi-classical approach does not directly consider the target final state.  Multi-photon exchange, as in  Fig. \ref{fig:multiphoton} can be included by adding an impact-parameter dependent additional-photon-exchange breakup probability to the cross-section calculation \cite{Baltz:2002pp}.  The energy scale for nuclear excitation is much lower than for vector meson production.  So, over the time scale required for vector meson production,
the target does not lose significant energy, and the excitation should not change the $\vec{x}_i$ distribution significantly. 
 
Equation \ref{eq:sigmasimp} can be applied to $J/\psi$ photoproduction in peripheral collisions, albeit with some open questions \cite{Zha:2017jch,Klusek-Gawenda:2015hja}: Does photoproduction involve the participant (in the hadronic interaction) nucleons, or only the spectators?  Hadroproduction has a similar (or shorter, depending on the interaction) time scale than photoproduction.  Does the hadronic interaction occur before the photoproduction?  If so, the participant nucleons will have a lower energy, reducing the photonuclear cross-section.   A full calculation should consider both time orderings.  This is possible by expanding Eq. \ref{eq:sigmasimp}, but it is incompatible with the GW approach.

In the GW paradigm, factorization (Fig. \ref{fig:multiphoton}) fails.   Different fluctuations of the incident photon interact with the nuclear target with different cross-sections.  If the target is excited, though, it also fluctuates.   $A(K,\Omega)$ in Eqs. \ref{eq:tot} through \ref{eq:inc} would need to be expanded to include an additional dependence on the final state configuration $\Omega'$: $A(K,\Omega,\Omega')$.  To get the total cross-section, it would then be necessary to sum over the different final state configurations.  The only simple division into coherent and incoherent production would be the one expected in GW: coherent production requres that $\Omega' = \Omega$.  Any other division including times scales, {\it etc.} would depend on the final state configuration in a complex manner.

The GW paradigm uses the optical model and assumes that the excitation time is short compared to the time required to propagate through the nucleus \cite{Miettinen:1978xv,Caneschi:1977zt}.  Because 'Diffraction scattering arises as the shadow of nondiffractive multiparticle production.'  propagation-time effects that pertain to non-diffractive multiparticle production can also affect diffractive production.  So, the dependence on final state excitation is not simple.  

There are, of course, some complications to either approach.  For lighter mesons it is necessary to account for the possibility that the $q\overline q$ dipole interacts multiple times as it passes through the nucleus.  This is usually done with a Glauber calculation \cite{Gribov:1968jf}. These corrections reduce the cross-section, but do not alter the arguments presented here.  Glauber calculations reproduce the observed cross-sections for photoproduction of light vector mesons on heavy nuclei at low $p_T$ (in the coherent regime), either with or without excitation  \cite{Klein:1999qj,Frankfurt:2002sv}; the agreement improves by including excited intermediate states of the dipole \cite{Frankfurt:2015cwa}.  For heavy vector mesons, gluon shadowing must be included \cite{ALICE:2021gpt, Guzey:2013xba}.  

Equation \ref{eq:sigmasimp} could be expanded to include the partonic structure of a nucleus, by changing the sum over $i$ to run over the quarks in each nucleon and replacing dipole-nucleon cross-section and couplings with their quark counterparts.  The number of quarks depends on the lowest kinematically allowable Bjorken$-x$, evaluated at a $Q^2$ corresponding to half of the vector meson mass ($M_V$): $x_{\rm min} = M_V^2/2km_p$, where $M_V$ and $m_p$ are the vector meson and proton masses, and here $k$ is in the target frame.  As $k$ rises, $x_{\rm min}$ drops, and the number of energetically accessible quark targets increases.   Since the intra-quark separation is smaller than the intra-nucleon separation, partonic interactions add a third spectral component, corresponding to nucleon breakup, with a harder $p_T$ spectrum than the incoherent component \cite{ALICE:2021gpt}. 

\section{Why Good-Walker fails}

GW does not explain coherent photoproduction with nuclear breakup or in hadronic collisions because these reactions do not satisfy a key GW assumption: that the input projectile is a single photon (or other particle).  In UPCs, as Fig. \ref{fig:multiphoton} shows, the electromagnetic fields are strong enough so that multiple photons can be exchanged, violating the GW assumption of a single incident particle.  The presence of hadronically interacting particles in peripheral collisions likewise violates this assumption.   The possible presence of additional photons is enough to violate GW, even if they (or their effects) are not observed.  Reactions with a second photon effectively absorb some of the expected one-photon cross-section.   They also reduce the energy of the incident electron or ion, contribute to the measured $t$, and  can alter the apparent division between coherent and incoherent cross-sections.   In fact, it is impossible to tell if a reaction that produces a vector meson and excites a nucleus occurs via one-photon or two-photon exchange.  These channels can interfere with each other; at  $p_T\approx 5 \hbar/R_A$ the amplitudes for the one-photon and two-photon processes are similar. 

Instead of taking the incident particle is a single photon, the incident particle could be the electron/proton/ion.   This would invalidate the single-photon explanation, but
does not solve the problem that coherence is visible even when the scattering is inelastic because the target (or the projectile) breaks up in the reaction.  

There are other possible complications to the reaction.  The nuclear target could radiate an unseen bremsstrahlung photon, carrying off energy and momentum.  Bremsstrahlung is infrared divergent, so, for a low enough cutoff energy, radiation is present in every interaction.  

The photon emitter and nuclear target can also each emit a photon, which fuse to produce a lepton pair.  The cross-section for lepton pair production is finite because of the lepton masses, but it is large.  For lead-lead collisions at the LHC, the cross-section for two-photon production of $e^+e^-$ pairs is about 200,000 barns \cite {Baur:2007zz,Baltz:2007kq}. For near-grazing lead-lead collisions ($b\approx 2R_A$), the average number of produced pairs is more than one \cite{Alscher:1996mja}.

Most of these pairs are invisible to RHIC or LHC detectors, since the leptons have a low $p_T$ and/or a high rapidity.  These soft particles are unlikely to affect the overall kinematics of the vector meson production.  However, they do break the GW paradigm, in which the initial and final state of the target should be the same.  

The likelihood of multi-photon exchange is much higher for heavy-ion photon-sources than for proton or electron radiators, but this is a question of degree, rather than a qualitative difference.  Two-photon exchanges have been seen to have implications for elastic form factor measurements in $eA$ collisions, for example \cite{Blunden:2003sp,Arrington:2007ux}.  

Theoretical underpinnings aside, both GW and Eq. \ref{eq:sigmasimp} make similar predictions for coherent production.   Since $p_T$ is conjugate to the impact parameter $b$ and $t\approx p_T^2$, in both approaches the two-dimensional Fourier transform of $\sqrt{d\sigma_{\rm coh}/dt}$ gives, with some caveats, the transverse profile of interaction sites in the target \cite{Munier:2001nr,Diehl:2003ny,Toll:2012mb,Klein:2019qfb,Klein:2021mgd}.   

However, for incoherent production, the implications are significant. Equation \ref{eq:inc} does not have a counterpart in the semi-classical approach; there is no association between event-by-event fluctuations and the incoherent cross-section.  Instead, the incoherent cross-section depends on the number of emitters, with the $p_T$ spectrum of the incoherent production depending on the sizes of the individual emitters.  A nucleus consisting of fixed, static nucleons would still interact incoherently.  In contrast, in GW,  without fluctuations, the incoherent cross-section is zero.  

The semiclassical approach again finds support in the STAR $\rho$ photoproduction data, where the incoherent cross-section at large $|t|$ ($0.2 < |t|<0.45$ GeV$^2$, where the coherent cross-section is small) was fit to a dipole form factor, consistent with a single proton target  \cite{STAR:2017enh,Klein:2021mgd}.  The form is also seen in color glass condensate calculations \cite{Mantysaari:2022sux}. In contrast, an exponential function gave a poor fit to the data.   In GW, the incoherent cross-section is driven by fluctuations, and there is no reason to expect it to follow a dipole form factor. 

\section{Other issues with $d\sigma_{\rm inc}/dt$ at low $|t|$}

There are some other problematic issues with common treatments of $d\sigma_{\rm inc}/dt$ at low $|t|$.  Some of these difficulties further challenge the GW approach.  

One issue arises at very small $|t|$, where energy conservation limits incoherent interactions. Nuclear excitation is endothermic.   As $|t|$ decreases, the energy transfer to the nucleus decreases, and, as $|t|\rightarrow 0$ there is insufficient energy transferred to excite the nucleus, so incoherent interactions become impossible.  This problem is even worse for a dipole form factor, because $d\sigma_{\rm inc}/dt$ rises as $|t|$ decreases, even as $|t|\rightarrow 0$. 

The decrease as $|t|\rightarrow 0$ is seen in the semiclassical approach.  In Eq. \ref{eq:sigmasimp} as $|t|\rightarrow 0$,  the coherent cross-section absorbs all of the available amplitude, squeezing out incoherent production.   Under the assumption of randomly positioned static nucleons, it is possible to quantify the degree of incoherence via the deviation of $\exp(i\vec{k}\cdot\vec{x}_i)$ from one.   Expanding the exponential as a Taylor series yields
$d\sigma/dk \propto k^2$;
it drops quadratically as $k$ decreases.  This approach neglects several factors, including the nucleon-nucleon repulsive force (which creates correlations in the nucleon positions) and the relatively sharp nuclear edges, but does demonstrate the asymptotic behavior. 

Nuclei can dissociate in many ways, depending on the available energy \cite{Chang:2021jnu}. Common modes are via  neutron emission (via a Giant Dipole Resonance (GDR) or other intermediate excited state), proton emission, or photon emission.    Although low-energy nuclear excitations like the GDR are often considered collective oscillations, they can also be described in terms of single-particle transitions \cite{Brink}, so may be produced by a Pomeron interacting with a single nucleon. 

Table \ref{tab:nuclei} shows that the energies required to eject nucleons from two commonly accelerated nuclei, $^{197}$Au and $^{208}$Pb, range from 5.27 to 8.07 MeV.  This energy is required to break up the bound nuclei.  Additional energy can further excite the target, or may provide kinetic energy to the ejected nucleon.

If the Pomeron transfers its energy to a single nucleon, as indicated by the STAR data \cite{STAR:2017enh},
these thresholds correspond to a minimum initial nucleon recoil momentum of 100 to 125 MeV/c.  The struck nucleon will transfer energy to the target, but this sets the scale for the Pomeron $p_T$.   At substantially smaller Pomeron $p_T$, only excitation followed by photon emission is possible.  In this low-$p_T$ region $d\sigma_{\rm incoherent}/dt$ may be substantially smaller than an extrapolation from higher $|t|$ would indicate. 

\begin{table*}
\begin{tabular}{|c c | c c|}
\hline
Lead  &  Mass & Gold & Mass  \\
\hline
$^{208}$Pb &   207.976627 Dal & $^{197}$Au & 196.966569 Dal \\
$^{207}$Pb & 206.975872 Dal  & $^{196}$Au &195.96657 Dal \\
$\Delta E$ (n emiss.) & 7.38 MeV & $\Delta E$ (n emiss. ) & 8.07 MeV \\
\hline
$^{207}$Tl &  206.975872   & $^{196}$Pt & 195.964952 Dal \\
$\Delta E$ (p emiss.) & 7.57 MeV & $\Delta E$ (p emiss.) & 5.27 MeV \\
\hline
\end{tabular}
\label{tab:nuclei}
\caption{The masses (in Daltons) for $^{208}$Pb  and $^{197}$Au for the remnant nuclei after 1n and 1p emission \cite{masses}.   All reactions are endothermic, and the $\Delta E$s are the energy in MeV required for the reaction to proceed.}
\end{table*}

Lower energy excitations come from transitions between nuclear shell-model states, at specific excitation energies. The energy spectra are very different for the two commonly used heavy nuclei \cite{levels}. $^{208}$Pb is doubly magic, so is very stable, with a lowest lying excited state at 2.6 MeV.  In contrast, $^{197}$Au has its lowest lying state at 77 keV.  This state has a 1.9 nsec lifetime, so, for a 110 GeV/n gold nucleus (expected at the EIC), the characteristic decay distance is about 70 m, long enough so that the excited nucleus will decay far outside any realistic detector.  This excitation is essentially invisible.  The next excited states are at 269 and 279 keV.

The different excitation energy spectra have significant consequences for the GW paradigm.  Both $^{197}$Au and $^{208}$Pb, have similar sizes, and their density distributions are both well described by a Woods-Saxon distributions. They should have similar distributions of low$-x$ gluonic hotspots.  In GW, the incoherent photoproduction cross-sections at small $|t|$ should be quite similar.   But, their different shell-model excitation spectra should lead to differences in 
$d\sigma_{\rm inc}/dt$ at low $|t|$. 

To apply GW, it is necessary to accurately classify reactions as coherent or incoherent.  No present or planned RHIC or LHC detector could detect these de-excitation photons in UPCs.  EIC  detector collaborations are planning ZDCs that can detect low-energy photons, but their energy threshold will be limited by the synchrotron radiation background, which is likely to be too high to permit the observation of photons from gold de-excitation \cite{AbdulKhalek:2021gbh}. Even with lead targets, some of the photons are emitted opposite to the direction of motion, so will be Lorentz downshifted, rather than boosted.  This will limit the overall detection efficiency.

Although lead is preferred over gold, it is not a panacea. The relative excitation probabilities to different excited states due to Pomeron exchange are not well known, so the relative excitation probabilities to different states are not well known, so determining detection efficiency will be difficult.  Different Monte Carlo codes have taken rather different approaches. 

STARlight (for UPCs) \cite{Klein:2016yzr} and eSTARlight (for ep/eA collisions) \cite{Lomnitz:2018juf} both largely follow Eq. \ref{eq:sigmasimp}, using nuclear and nucleon form factors to predict the $p_T$ spectra for coherent and incoherent production respectively.  However, they do not model the depletion of incoherent production at small $p_T$.  STARlight has been shown to provide a good description of light vector meson photoproduction cross-sections \cite{STAR:2007elq, ALICE:2021jnv,ALICE:2020ugp}, although for the $J/\psi$, it overpredicts the cross-section \cite{ALICE:2021gpt,CMS:2016itn}, likely because it does not include gluon shadowing. 

The Sar{\it t}re generator \cite{Toll:2012mb,Toll:2013gda} follows the GW approach.  For each reaction studied, it generates 500 random nuclear configurations, and then calculates the coherent and incoherent $d\sigma/dt$ using Eqs. \ref{eq:coh} and \ref{eq:inc}, using a dipole model approach.    In Sar{\it t}re, $d\sigma_{\rm coh}/dt$ roughly follows an exponential behavior at large $|t|$, but exhibits a small downturn for $|t|<0.015$ GeV$^2$.  The downturn reduces the cross-section for incoherent $\phi$ electroproduction a factor of 2-3 as $|t|\rightarrow 0$ compared to the exponential baseline.  This is a smaller reduction than the Taylor expansion of $d\sigma/dk$ predicts.  The calculated cross-section also lacks structure due to nuclear levels. Instead, the nuclear excitation energies $E$ are chosen to follow a $1/E^2$ distribution, with the nuclear breakup being done by a statistical modelling code. 

Benchmark eA Generator for LEptoproduction (BeAGLE) is a Monte Carlo code that simulates a variety of $ep$ and $eA$ collisions, including incoherent vector meson production \cite{Chang:2022hkt}.  Interactions involve a randomly chosen nucleon target, producing hadrons; the nuclear recoil is simulated with a cascade model, with FLUKA handling the low-energy nuclear remnants.   BeAGLE makes similar predictions about $d\sigma/dt$ as SARTRE for incoherent $J/\psi$ photoproduction. 

\section{Conclusions}

In conclusion, we have shown that the Good-Walker paradigm fails to explain two classes of events that have been observed in relativistic heavy-ion collisions:  coherent photoproduction accompanied by mutual Coulomb excitation in UPCs, and coherent photoproduction in peripheral heavy-ion collisions.   A semi-classical approach based on adding amplitudes is much more effective in these cases.  These two approaches make similar predictions for coherent production, but have very different takes on incoherent production.     Reconciling these two approaches is critical for understanding how incoherent production is sensitive to fluctuations in the average nuclear configuration.     

Because incoherent photoproduction involves nuclear breakup, it is an exothermic process.  As $|t|$ decreases, some breakup channels will become energetically inaccessible, so $d\sigma_{\rm inc}/dt$ is unlikely to be a single smooth curve. 
 
Looking ahead, it is important to improve calculations to better quantify the effect of multi-photon exchange, bremsstrahlung and pair production on the division into coherent and incoherent production, especially for ep/eA collisions.    This might involve applying a quantum field theory approach to the GW paradigm.  Although existing studies of HERA data on proton targets may not be significantly affected, future high-precision studies are likely to reach its limits. 

It is also important to develop better models of nuclear excitation due to Pomeron exchange.  Studies of photoexcitation \cite{Pshenichnov:2001qd} are clearly relevant here, but may not transfer 100\% since the Pomeron couples equally to protons and neutrons. These models will be needed to correct data for mis-classification of coherent and incoherent photoproduction due to unobserved breakup. 

It may be possible to collect relevant data at RHIC or the LHC, by installing a small forward electromagnetic calorimeter to detect photons from nuclear excitation.  In addition to observing nuclear excitations accompanying vector mesons, it can also study reactions involving only nuclear excitation \cite{Dmitrieva:2020ljw}.
When the EIC begins operations, much better data will become available.  Unfortunately, this is too late to inform the EIC detector designs. 

I  thank  Jesus Guillermo Contreras Nuno, Daniel Tapia-Takaki, Mark Strikman, Heikki Mantysaari, Igor Pshenichnov, Lee Bernstein and the LBNL Nuclear Structure group and  for useful discussions.  This work is supported in part by the U.S. Department of Energy, Office of Science, Office of Nuclear Physics, under contract numbers DE-AC02-05CH11231.  

\bibliographystyle{apsrev4-1} 
\bibliography{main}
\end{document}